\documentclass[aps,superscriptaddress,twocolumn,pra]{revtex4}

\usepackage{graphicx}
\begin{document}

\title{Transport and localization of waves in ladder-shaped lattices with locally $\mathcal{PT}$-symmetric potentials}

\author{Ba Phi Nguyen}
\email{nguyenbaphi@muce.edu.vn}
\affiliation{Department of Basic Sciences, Mientrung University of Civil Engineering, 24 Nguyen Du, Tuy Hoa, Vietnam}
\affiliation{Institute of NT-IT Fusion Technology, Ajou University, Suwon 16499, Korea}
\author{Kihong Kim}
\email{khkim@ajou.ac.kr}
\affiliation{Department of Energy Systems Research and Department of Physics, Ajou University, Suwon 16499, Korea}

\begin{abstract}
We study numerically the transport and localization properties of waves in ordered and disordered ladder-shaped lattices with local $\mathcal{PT}$ symmetry. Using a transfer matrix method, we calculate the transmittance and the reflectance for the individual channels and the Lyapunov exponent for the whole system. In the absence of disorder, we find that when the gain/loss parameter $\rho$ is smaller than the interchain coupling parameter $t_{v}$, the transmittance and the reflectance are periodic functions of the system size, whereas when $\rho$ is larger than $t_{v}$, the transmittance is found to be an exponentially-decaying function while the reflectance attains a saturation value in the thermodynamic limit. For a fixed system size, there appear perfect transmission resonances in each individual channel at several values of the gain/loss strength smaller than $t_{v}$. A singular behavior of the transmittance is also found to appear at various values of $\rho$ for a given system size. When disorder is inserted into the on-site potentials, these behaviors are changed substantially due to the interplay between disorder and the gain/loss effect. When $\rho$ is smaller than $t_{v}$, we find that the presence of locally $\mathcal{PT}$-symmetric potentials suppresses Anderson localization, as compared to the localization in the corresponding Hermitian system. When $\rho$ is larger than $t_{v}$, we find that  localization becomes more pronounced at higher gain/loss strengths. We also find that the phenomenon of anomalous localization occurs in disordered locally $\mathcal{PT}$-symmetric systems precisely at the spectral positions $E=0$ and $E=\pm\sqrt{t_{v}^2-\rho^2}$. The anomaly at the band center manifests as a sharp peak contrary to the conventional cases, whereas the anomalies at $E=\pm\sqrt{t_{v}^2-\rho^2}$ manifest as sharp dips.
\end{abstract}

\maketitle

\section{Introduction}

Anderson localization, which is one of the most fundamental physical phenomena in many areas of physics, occurs due to the interference of wave components multiply scattered by randomly placed scattering centers \cite{And,Lee,She}. It has been observed experimentally in many different kinds of waves including microwaves \cite{Dal,Cha}, optical waves \cite{Sch,Lah} and matter waves \cite{Bil,Roa}. Anderson localization has been studied mainly in conservative, Hermitian systems, as it is commonly assumed that disorder-induced localization is caused by multiple scattering from the real part of the potential. In recent years, however, considerable attention has been paid to localization phenomena occurring in non-Hermitian systems \cite{Hat,Asa,Sil,Jov,Kal,Eic,Bas,Vaz,Mej,Kar}. An example of this kind of systems is the Hatano-Nelson model, where the presence of a constant imaginary vector potential in the Anderson Hamiltonian gives rise to the transition from a real to a complex spectrum, which is associated with the existence of a mobility edge \cite{Hat}.

In recent years, there has been significant attention devoted to physical systems which do not obey parity ($\mathcal{P}$) and time-reversal ($\mathcal{T}$) symmetries separately but exhibit a combined parity-time ($\mathcal{PT}$) symmetry. A seminal idea for such a system was first proposed by Bender and co-workers, who demonstrated that a class of non-Hermitian Hamiltonians could possess entirely real eigenvalue spectra  \cite{Ben1,Ben2}. In this context, $\mathcal{PT}$ symmetry requires that the imaginary part of the complex potential $V$ in the Hamiltonian is an antisymmetric function of the position $\bf r$, whereas its real part is a symmetric one, therefore $V({\bf{r}})=V^{*}(-{\bf{r}})$. The eigenvalue spectrum remains real and $\mathcal{PT}$ symmetry is unbroken, if the magnitude of the imaginary part of the potential does not exceed a threshold value. Above the threshold, the eigenvalue spectrum becomes complex and $\mathcal{PT}$ symmetry is said to be broken. For such systems, many intriguing phenomena have been predicted and observed in various fields of physics including solid state physics, atomic physics, optics and  metamaterials \cite{Jog,Elg,Mus,Mak,Guo,Bend1,Mir,Ben,Rut,Reg,Laz}.

The interplay between Anderson localization and $\mathcal{PT}$ symmetry is expected to open up an exciting new area of research. Even though many researches have been done on disordered $\mathcal{PT}$-symmetric systems, the majority of those have focused on the study of the spectral properties of the corresponding $\mathcal{PT}$-symmetric Hamiltonians, with less attention to their transport properties. In a recent paper, through an investigation of the transport properties of waves in one-dimensional (1D) randomly layered media with $\mathcal{PT}$ symmetry, the authors have found that the inverse localization length for such a system is the sum of the inverse localization length of a passive disordered medium and the inverse absorption/amplification length of an ordered $\mathcal{PT}$-symmetric medium \cite{Kal}. In another paper, it has been pointed out that the periodic transverse modulation of losses affects the expansion rate of localized excitations considerably \cite{Eic}. More recently, it has been found that for a disordered 1D binary optical lattice, the presence of gain and loss tends to favor the transport of extended excitations, while it inhibits that of localized excitations \cite{Mej}. By studying the propagation of light along the axis of an off-diagonal disordered optical waveguide array created by a simultaneous transverse modulation of the refractive index and the gain/loss parameter, Kartashov {\it et al.} have found that disorder-induced localization is suppressed when the system has unbroken $\mathcal{PT}$ symmetry \cite{Kar}. Beyond the symmetry-breaking threshold, the localization is restored and becomes stronger again deeply inside the regime of broken symmetry. In addition, in a uniformly disordered 2D complex lattice, the study of transverse localization of light has resulted in the conclusion that Anderson localization is enhanced in a $\mathcal{PT}$-symmetric potential \cite{Jov}.

In the presence of disorder, it has been shown that the $\mathcal{PT}$-symmetry-breaking threshold decreases as $L^{-2}$ in 1D disordered $\mathcal{PT}$-symmetric lattices, where $L$ is the system size \cite{Bend1}. In other words, the parameter region corresponding to an entirely real spectrum shrinks to zero in the thermodynamic limit, $L \to \infty$. On the other hand, Bendix {\it et al.} have proposed a novel class of disordered $\mathcal{PT}$ Hamiltonians, the spectra of which are entirely real in a broad range of parameters \cite{Bend2}. This kind of systems does not obey  {\it global} $\mathcal{PT}$ symmetry as usual, but possesses a symmetry termed {\it local} $\mathcal{PT}$ symmetry. In the absence of disorder, such locally $\mathcal{PT}$-symmetric structures have also been investigated in some recent papers \cite{Suc1,Suc2,Dam}. However, the effect of disorder on the transport properties of such structures is not fully understood. This provides us with the motivation to study research problems such as understanding how the transport properties are changed when the $\mathcal{PT}$-symmetry-breaking threshold is crossed in the presence of disorder.

In this paper, we present a numerical study of the transport and localization properties of waves in ordered and disordered locally $\mathcal{PT}$-symmetric systems. We employ the transfer matrix method developed by Heinrichs to calculate the transmittance, the reflectance and the Lyapunov exponent \cite{Hei}. In the absence of disorder, we find that when the gain/loss parameter $\rho$ is smaller than the interchain coupling parameter $t_{v}$, the transmittance and the reflectance are periodic functions of the system size. They have different dependencies on even and odd values of the system size. For a fixed system size, there appear transmission resonances in each individual channel at several values of the gain/loss strength. When $\rho$ is larger than $t_{v}$, we find that both the transmittance and the reflectance initially increase up to maximum values via oscillations and after that, the transmittance decays exponentially while the reflectance attains a saturation value as the system size increases to infinity. In addition, a singular behavior of the transmittance is found to appear at various values of $\rho$ for a given system size. As disorder is introduced in the on-site potential, these behaviors are changed substantially due to the interplay between disorder and the gain/loss effect. Specifically, when $\rho$ is smaller than $t_{v}$, we find that the presence of locally $\mathcal{PT}$-symmetric potentials suppresses Anderson localization, as compared to the localization in the corresponding Hermitian system. When $\rho$ is larger than $t_{v}$, we find that  localization becomes more pronounced at higher gain/loss strengths. This is quite consistent with the result presented in a recent paper \cite{Kar}. Finally, the phenomenon of anomalous localization is found to occur in the system under consideration at the special spectral positions with energy $E=0$ and $E=\pm\sqrt{t_{v}^2-\rho^2}$. Remarkably, the anomaly at the band center manifests as a sharp peak contrary to the conventional cases, whereas the anomalies at $E=\pm\sqrt{t_{v}^2-\rho^2}$ manifest as sharp dips.

The rest of this paper is organized as follows. In the next section, we introduce the theoretical model and describe the method of numerical calculation and the physical quantities of interest. In Sec.~III, we present our numerical results and discussions. Finally, we conclude the paper in Sec.~IV.

\section{Theoretical Model and Method}
\subsection{Model}

\begin{figure}
\includegraphics[width=7cm]{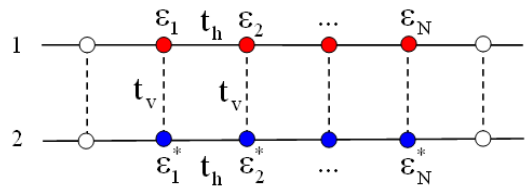}
\caption{Schematic view of the ladder-shaped lattice model consisting of two parallel chains of $N$ sites each. To study transport properties through the system, semi-infinite perfect leads are connected at both ends. $t_{h}$ and $t_{v}$ are the intrachain and interchain coupling parameters respectively and $\epsilon_n$ and $\epsilon_n^\ast$ are the on-site potentials.}
\end{figure}

Let us consider a ladder-structured lattice model consisting of two parallel chains of $N$ sites each, as shown in Fig.~1. If we consider only nearest-neighbor interactions, our model is governed by two coupled discrete Schr\"{o}dinger equations, which take the form
\begin{eqnarray}
t_{h}(\psi_{n-1,1}+\psi_{n+1,1})+t_{v}\psi_{n2}=(E-\epsilon_{n1})\psi_{n1},\nonumber \\
t_{h}(\psi_{n-1,2}+\psi_{n+1,2})+t_{v}\psi_{n1}=(E-\epsilon_{n2})\psi_{n2},
\label{equation1}
\end{eqnarray}
where $\psi_{nj}$ and $\epsilon_{nj}$ are the wave function amplitude and the on-site potential at site $n$ of chain $j$  $(j=1, 2)$. $t_{h}$ ($>0$) is the intrachain coupling parameter between the nearest-neighbor sites along the individual chains, while $t_{v}$ ($>0$) is the interchain coupling parameter connecting the $n$th sites of chains $1$ and $2$. This model is very similar to the one studied in Ref.~33, the only difference being that semi-infinite perfect leads are connected at both ends in the present model. We refer to it as unbounded model.

In order to study the combined effects of the simultaneous presence of disorder and $\mathcal{PT}$ symmetry on wave propagation in the system under consideration, we set $\epsilon_{n1}=\epsilon_{n}$ and $\epsilon_{n2}=\epsilon_{n}^{\ast}$ with $\epsilon_{n}=\beta_{n}+i\rho_{n}$, where $\beta_{n}$ is taken from a uniform random distribution in the range $[-\beta/2,\beta/2]$ and $\rho_{n}$ ($=\rho$) is a constant. For such a choice of the on-site potentials, the system possesses a local $\mathcal{PT}$ symmetry associated with each individual dimer enclosing one site $n$ of each chain \cite{Bend2}. Despite the lack of global $\mathcal{PT}$ symmetry, it has a parameter region corresponding to an exact $\mathcal{PT}$ phase in which the eigenvalues are real. To calculate the transfer and scattering matrices which are the basic elements of the transfer matrix method we have employed, it is convenient to cast Eq.~(\ref{equation1}) in the matrix form
\begin{eqnarray}
&&\left( \begin{array}{ccc}
\psi_{n-1,1}+\psi_{n+1,1} \\
\psi_{n-1,2}+\psi_{n+1,2}
\end{array} \right)\nonumber\\
&&~~~~=
\left( \begin{array}{ccc}
\left(E-\epsilon_{n}\right)/t_{h} & -t_{v}/t_{h} \\
-t_{v}/t_{h} & \left(E-\epsilon_{n}^{\ast}\right)/t_{h}
\end{array} \right)
\left( \begin{array}{ccc}
\psi_{n1} \\
\psi_{n2}
\end{array} \right).
\label{equation2}
\end{eqnarray}

In the perfect lead regions with zero on-site potentials, it is straightforward to diagonalize Eq.~(\ref{equation2})
by introducing two new {\it channel} wave functions $\phi_{n1}$ and $\phi_{n2}$ defined by
\begin{eqnarray}
\phi_{n1}=\frac{1}{\sqrt{2}}\left(\psi_{n1}+\psi_{n2}\right),~~
\phi_{n2}=\frac{1}{\sqrt{2}}\left(\psi_{n1}-\psi_{n2}\right)
\end{eqnarray}
and obtain the dispersion relations
\begin{eqnarray}
2 \cos k_{1}=\left(E-t_{v}\right)/t_{h},~2 \cos k_{2}=\left(E+t_{v}\right)/t_{h},
\label{equation3}
\end{eqnarray}
where $k_{1}$ and $k_{2}$ are the wave numbers associated with the channels 1 and 2 respectively. Let us assume that $0<t_{v}/t_{h}<2$. Then both channels are propagating when $-2+t_{v}/t_{h}<E/t_{h}<2-t_{v}/t_{h}$, while one channel is propagating and the other is evanescent when $-2-t_{v}/t_{h}<E/t_{h}<-2+t_{v}/t_{h}$ or $2-t_{v}/t_{h}<E/t_{h}<2+t_{v}/t_{h}$.

It is important to notice that there is a fundamental difference between the eigenvalue problem for the bounded model studied in Ref.~33 and the scattering problem for the unbounded model studied in this paper. In the former, the authors fixed the gain/loss parameter $\rho$ and the boundary conditions and obtained the corresponding energy eigenvalues, which are in general complex-valued. In contrast, $E$ in the scattering problem is the energy of the incident wave and is a real-valued free parameter.

\subsection{Method}

In this paper, we will study the transmission and reflection properties of both ordered and disordered locally $\mathcal{PT}$-symmetric systems. In particular, we are interested in calculating the $2\times 2$ matrix transmittance $T_{ij}$ and reflectance $R_{ij}$, where the index $i$ refers to the reflected and transmitted channels and $j$ refers to the incident channel, and the Lyapunov exponent $\gamma$. The latter quantity is equal to the inverse of the localization length $\xi$ and is defined by
\begin{eqnarray}
\gamma(E)=\frac{1}{\xi(E)}=-\lim_{L \to \infty} \frac{\langle \ln g\rangle}{2L},
\label{equation4}
\end{eqnarray}
where $L$ is the system size measured in the unit of the lattice spacing and is equal to $N$ and $g$ is the dimensionless conductance. This quantity is used
to distinguish between localized and extended states. The angular bracket $\langle\cdots\rangle$ stands for averaging over
a large number of distinct
disorder configurations.

In the context of electron transport in mesoscopic systems, the conductance $G$ and the dimensionless conductance $g$
in the zero-temperature limit are related to the transmission probability via the Landauer two-probe formula:
\begin{eqnarray}
g=\frac{h}{2e^2}G={\rm Tr} \left(\hat{t}\hat{t}^{\dagger}\right),
\label{equation5}
\end{eqnarray}
where $e$ is the electron charge and $h$ is Planck's constant \cite{Imr}. The reflection matrix $\hat{r}$ and
the transmission matrix $\hat{t}$ are written as
\begin{eqnarray}
\hat{r}=
\left( \begin{array}{ccc}
r_{11} & r_{12}\\
r_{21} & r_{22}
\end{array} \right),~~
\hat{t}=
\left( \begin{array}{ccc}
t_{11} & t_{12}\\
t_{21} & t_{22}
\end{array} \right),
\label{equation6}
\end{eqnarray}
where $r_{ij}$ and $t_{ij}$ ($i,j=1,2$) denote the reflected and transmitted wave amplitudes in the $i$th channel
when there is a unit flux incident in the $j$th channel. These quantities are given by
\begin{eqnarray}
\hat{r}&=&
\frac{1}{\delta}\left( \begin{array}{ccc}
X_{12}X_{44}-X_{42}X_{14} & X_{22}X_{14}-X_{12}X_{24}\\
X_{32}X_{44}-X_{42}X_{34} & X_{22}X_{34}-X_{32}X_{24}
\end{array} \right),\nonumber\\
\hat{t}&=&
\frac{1}{\delta}\left( \begin{array}{ccc}
X_{44} & -X_{24}\\
-X_{42} & X_{22}
\end{array} \right),
\label{equation8}
\end{eqnarray}
where $\delta=X_{22}X_{44}-X_{24}X_{42}$. $X_{ij}$ ($i,j=1,2,3,4$) denotes the elements of the transfer matrix for the whole system of length $N$.
Details of the method of calculating $\hat{r}$ and $\hat{t}$ can be found in Ref.~37

The transmittance and the reflectance are determined by $T_{ij}=\vert t_{ij}\vert^2$ and $R_{ij}=\vert r_{ij}\vert^2$ respectively. Since our model is non-Hermitian, the conservation of probability current is generally not satisfied. For instance, for a unit flux incident in the second channel, one has $\vert r_{12}\vert ^2+\vert r_{22}\vert ^2+\vert t_{12}\vert ^2+\vert t_{22}\vert ^2\not=1$ in general.

\section{Results}
In our calculations, all the energies are measured in the unit of $t_{h}$, which we set equal to 1. We will restrict our attention to the situation where both of the two channels are propagating, therefore $-2+t_{v}/t_{h}<E/t_{h}<2-t_{v}/t_{h}$. In that case, the transport and localization
properties are qualitatively similar for both channels. In fact, they are identical in the symmetric case where $E=0$. In the majority of the calculations to
be presented in this paper, we will set $E=0$ and concentrate on the second channel.

\subsection{Ordered case}

We first consider the situation where no disorder in the on-site potential is present ($\beta=0$). In this case, the only requirement ensuring  unbroken $\mathcal{PT}$ symmetry in the corresponding bounded model with no semi-infinite perfect leads is that the gain/loss strength $\rho$ must be smaller than or equal to the interchain coupling strength $t_{v}$, namely $\rho\le \rho_{c}=t_{v}$ \cite{Bend2,Suc2}. This means that the $\mathcal{PT}$-symmetry-breaking threshold does not depend on the intrachain coupling parameter $t_{h}$. We fix the energy of the incident wave at $E=0$ at the band center for all the results obtained in this subsection. The specific results may be quantitatively different for different values of $E$, but the qualitative behaviors of the system are similar.

\begin{figure}
\includegraphics[width=9cm]{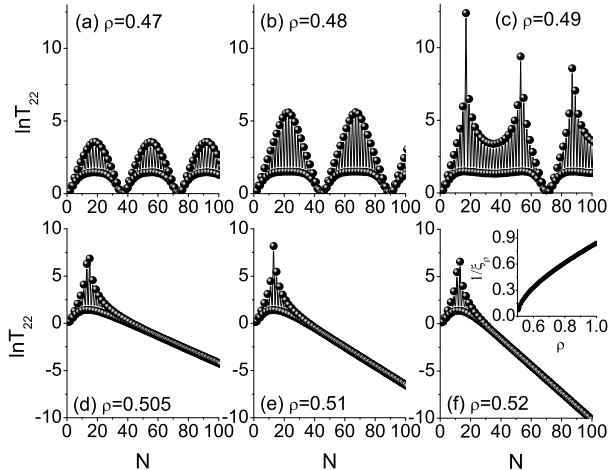}
\caption{$\ln T_{22}$ plotted versus system size $N$ for various values of the gain/loss parameter $\rho$, when the interchain coupling parameter $t_v$ is fixed to 0.5. In (a), (b) and (c), where $\rho$ is smaller than $t_v$, the transmittance depends periodically on $N$. In (d), (e) and (f), where $\rho$ is larger than $t_v$, the transmittance decays exponentially as $N$ becomes sufficiently large. In the inset of (f), the exponential decay rate of the transmittance, $1/\xi_\rho$, is plotted versus $\rho$ when $\rho>t_{v}$.}
\end{figure}
\begin{figure}
\includegraphics[width=9cm]{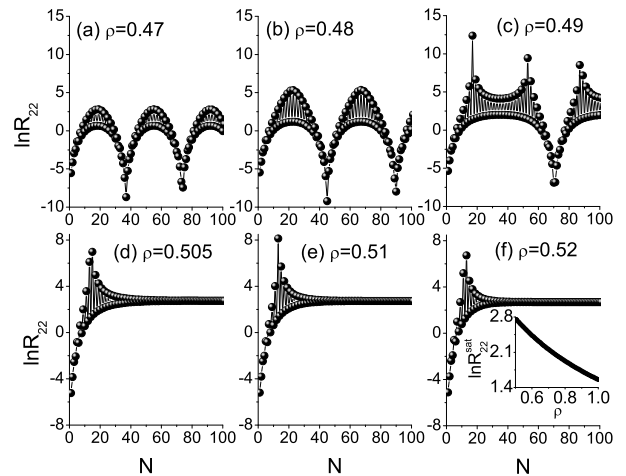}
\caption{$\ln R_{22}$ plotted versus system size $N$ for the same parameter values as in Fig.~2. In (a), (b) and (c), where $\rho$ is smaller than $t_v$ ($=0.5$), the reflectance depends periodically on $N$. In (d), (e) and (f), where $\rho$ is larger than $t_v$, the reflectance approaches a constant value as $N$ becomes sufficiently large. In the inset of (f), the saturation value of the logarithmic reflectance, $\ln R_{22}^{\rm sat}$, is plotted versus $\rho$ when $\rho>t_{v}$.}
\end{figure}

\begin{figure}
\includegraphics[width=9cm]{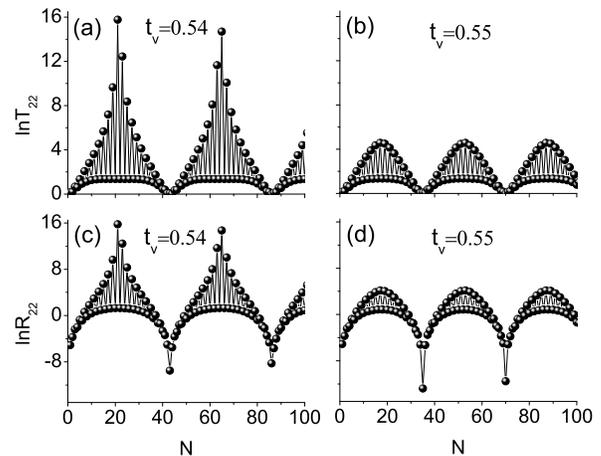}
\caption{$\ln T_{22}$ and $\ln R_{22}$ plotted versus system size $N$ for two different values of the interchain coupling parameter $t_{v}$, when the gain/loss parameter $\rho$ is fixed to 0.52. The transmittance and the reflectance depend on $N$ periodically because $\rho<t_{v}$.}
\end{figure}

In Figs.~2 and 3, we plot the logarithmic transmittance $\ln T_{22}$ and the logarithmic reflectance $\ln R_{22}$ versus system size $N$ for various values of the gain/loss parameter $\rho$, when $t_v$ is fixed to 0.5. We find that two distinct behaviors occur depending on the relative size of
$\rho$ with respect to the critical value, $\rho_{c}=0.5$. When $\rho$ is smaller than $\rho_c$, the transmittance and the reflectance are periodic functions of $N$. This period increases rapidly as $\rho$ increases and approaches infinity as $\rho$ goes to $\rho_c$. In this regime, apart from the set of resonances with $T_{22}=1$ (or $\ln T_{22}=0$) occurring at certain periodic values of $N$, the transmittance is always larger than 1. The number of such resonances depends on $N$ and $\rho$. For a given $N$, it decreases with increasing $\rho$ and vanishes as $\rho$ approaches $\rho_c$. At the same values of $N$ at which $T_{22}=1$, $R_{22}$ takes very small values close to zero. We note that a similar behavior has been found in a purely 1D $\mathcal{PT}$-symmetric tight-binding lattice model \cite{Vaz}. We also observe that both the transmittance and the reflectance behave
differently for even and odd values of $N$, their values for odd $N$ being substantially larger than those for neighboring even values of $N$. This is a characteristic feature of coherently amplifying/absorbing media and is not associated with the presence of local $\mathcal{PT}$ symmetry \cite{Jiang,Datta}.

When $\rho$ is greater than $\rho_c$, both the transmittance and the reflectance show completely different behaviors. In particular, it is found that there exists a certain value of the system size, $N_{\rm max}$, below which the transmittance and the reflectance grow
and above which they decay with increasing $N$ via oscillations which arise due to their different dependencies on even and odd values of $N$. Well above $N_{\rm max}$, the amplitudes of the oscillations decay away. The transmittance decreases exponentially and the reflectance approaches a constant value as $N$ increases to large values. The rate of the exponential decay of $T_{22}$, which is defined by
\begin{equation}
\frac{1}{\xi_\rho}= -\lim_{N\to \infty}\frac{\ln T_{22}}{N},
\end{equation}
increases as $\rho$ increases as shown in the inset of Fig.~2(f), while the saturation value of $R_{22}$ decreases monotonically as in Fig.~3(f).

The transition from a periodic behavior to either an exponential decay ($T_{22}$) or a saturation behavior ($R_{22}$) in the thermodynamic limit, when $\rho$ is increased across the critical value, is equivalent to a manifestation of the broken $\mathcal{PT}$ symmetry in the corresponding bounded model. In Fig.~4, we show the result obtained when $\rho$ is fixed and $t_v$ is varied. When $\rho$ is smaller than $t_{v}$, the dependence of the transmittance and the reflectance on $N$ is found to be always periodic. We also observe that when $\rho$ is very close to but smaller than
$t_v$, both $T_{22}$ and $R_{22}$ display sharp peaks at certain periodic values of $N$, as shown in Figs.~4(a) and 4(c) and also in Figs.~2(c)
and 3(c).

\begin{figure}
\includegraphics[width=9cm]{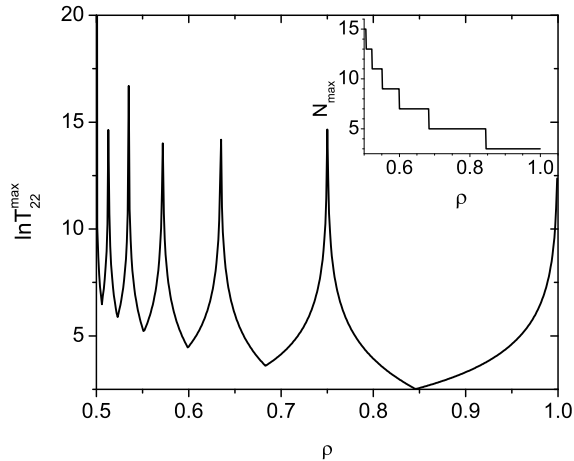}
\caption{$\ln T^{\rm max}_{22}$ plotted versus gain/loss parameter $\rho$, when the interchain coupling parameter $t_{v}$ is fixed to 0.5. The singular behavior of $T^{\rm max}_{22}$ appears at different values of $\rho$. In the inset, the system size at which the transmittance shows a maximum, $N_{\rm max}$, is plotted versus $\rho$.}
\end{figure}

\begin{figure}
\includegraphics[width=9cm]{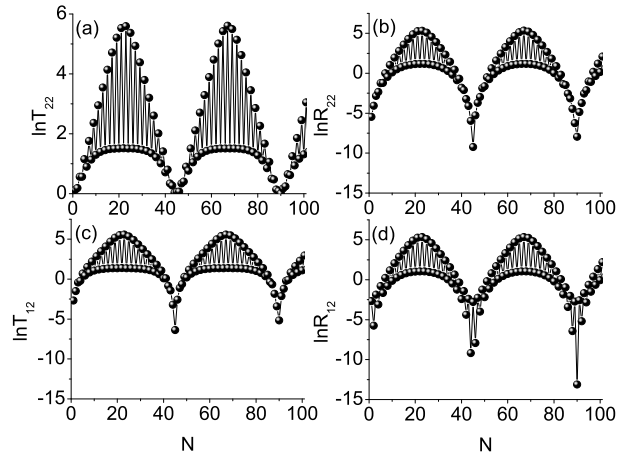}
\caption{$\ln T_{22}$, $\ln R_{22}$, $\ln T_{12}$ and $\ln R_{12}$ plotted versus system size $N$ when $t_{v}=0.5$ and $\rho=0.48$. At the system sizes for which $T_{22}=1$, which are $N=45$ and 90, all of $T_{12}$, $R_{12}$ and $R_{22}$ take very small values close to zero.}
\end{figure}

\begin{figure}
\includegraphics[width=8cm]{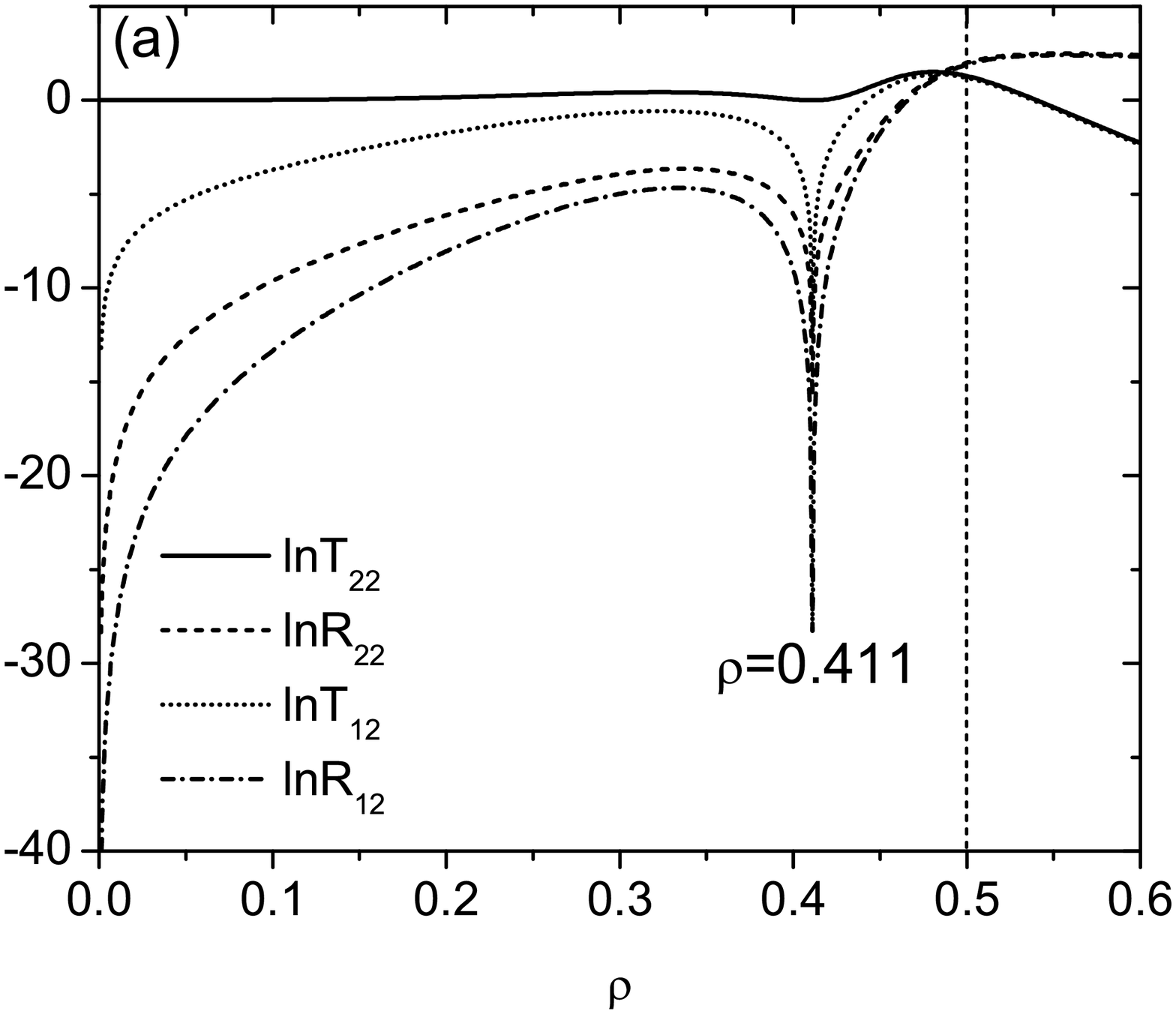}
\includegraphics[width=8cm]{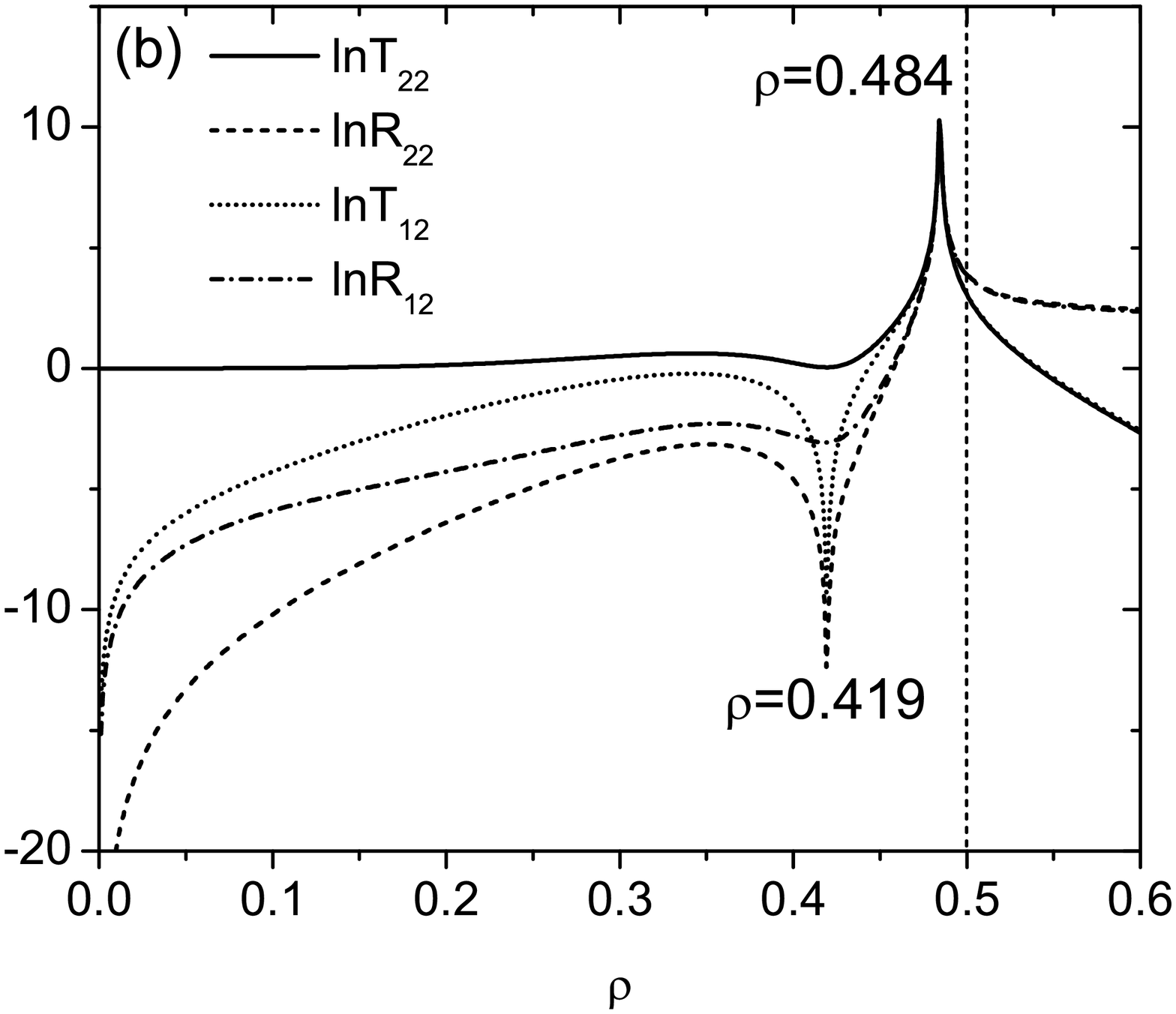}
\includegraphics[width=8cm]{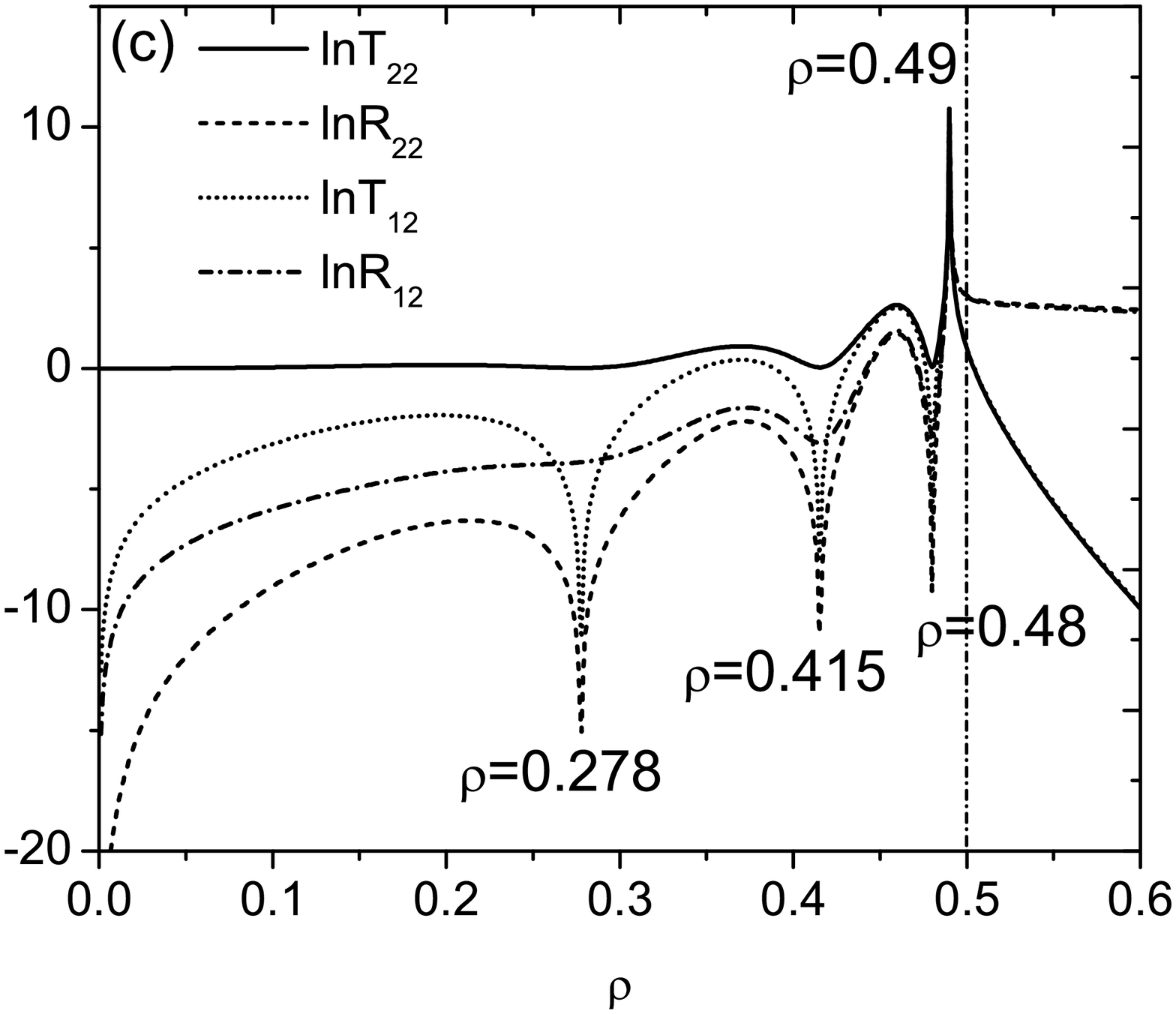}
\caption{$\ln T_{22}$, $\ln R_{22}$, $\ln T_{12}$ and $\ln R_{12}$ plotted versus gain/loss parameter $\rho$, when $t_{v}=0.5$ and (a) $N=22$, (b) $N=23$, (c) $N=45$. With proper choices of $N$ and $\rho$, the system with local $\mathcal{PT}$ symmetry can act as a passive one exhibiting perfect transmission resonances, where $T_{22}=1$ and $R_{22}=T_{12}=R_{12}=0$, in each individual channel. The values of $\rho$ at which the resonances occur are
designated in each figure, together with the value of $\rho$ at the peak where all of $T_{22}$, $R_{22}$, $T_{12}$ and $R_{12}$ take very large values. The number of the transmission resonances increases with $N$.}
\end{figure}

In the region where $\rho>\rho_c$, the transmittance and the reflectance take their maximum values at a certain system size $N_{\rm max}$, the value of which depends on $\rho$. In Fig.~5, we plot $\ln T^{\rm max}_{22}$ as a function of $\rho$. We find that there appear many sharp peaks indicating the presence of singularities at different values of $\rho$. This kind of singular behavior has been observed in 1D amplifying media without disorder \cite{Datta}. In the presence of disorder or nonlinearity in the system, these singularities are known to be destroyed \cite{Joshi,Ngu1}. The dependence of $N_{\rm max}$ on $\rho$ is shown in the inset of Fig.~5.

In the case where a wave is incident in the second channel, the transmittance and the reflectance in the first channel, $\ln T_{12}$ and $\ln R_{12}$, and those in the second channel, $\ln T_{22}$ and $\ln R_{22}$, are plotted versus $N$ in Fig.~6. It is seen clearly that at the system sizes for which $T_{22}=1$, which are $N=45$ and 90 in this figure, all of $T_{12}$, $R_{12}$ and $R_{22}$ take very small values. Therefore, the sum, $T_{12}+T_{22}+R_{12}+R_{22}$, is unity at these values of $N$. This implies that with a proper choice of $N$ and $\rho$, our active system with local $\mathcal{PT}$ symmetry can act as a passive one exhibiting perfect transmission resonances in the individual channels. This interesting phenomenon is demonstrated in more detail in Fig.~7, where $\ln T_{22}$, $\ln R_{22}$, $\ln T_{12}$ and $\ln R_{12}$ are plotted versus $\rho$, when $t_{v}=0.5$ and $N=22$, 23, 45. In the case of $N=22$, there is only one transmission resonance occurring at $\rho=0.411$. When $N$ is equal to 23, the transmission resonance occurs at $\rho=0.419$. In this case, we also observe a very large peak at $\rho=0.484$, where all of $T_{22}$, $R_{22}$, $T_{12}$ and $R_{12}$ take very large values. The number of transmission resonances increases with increasing $N$. This is illustrated in Fig.~7(c) corresponding to $N=45$, which shows three transmission resonances at $\rho=0.278$, 0.415 and 0.48 and a large peak at $\rho=0.49$. We note that the resonance at $\rho=0.48$ has already been mentioned in Fig.~2. In the region where $\rho>\rho_{c}$, $T_{22}$ and $T_{12}$ decrease rapidly as $\rho$ increases, while $R_{22}$ and $R_{12}$ approach constant values larger than 1.

\subsection{Disordered case}

\begin{table*}
	\caption{\label{tab:table3} Decay rates extracted numerically from the logarithmic transmittances corresponding to passive disordered ($1/\xi_{\beta}$), ordered locally $\mathcal{PT}$-symmetric ($1/\xi_{\rho}$) and disordered locally $\mathcal{PT}$-symmetric ($1/\xi_{\beta+\rho}$) cases.}
	\begin{ruledtabular}
		\begin{tabular}{ccccc}
			&($\beta,~\rho$)&$1/\xi_{\beta}$ &$1/\xi_{\rho}$&$1/\xi_{\beta+\rho}$\\
			\hline
			&(0.5,~0)&0.0056&-&- \\
			&(0,~0.501)&-&0.0316&- \\
			&(0.5,~0.501)&-&-&0.0366\\
			&(0,~0.502)&-&0.0448&-\\
			&(0.5,~0.502)&-&-&0.0498\\
			&(0.4,~0)&0.0036&-&- \\
			&(0,~0.503)&-&0.0549&- \\
			&(0.4,~0.503)&-&-&0.0581\\
		\end{tabular}
	\end{ruledtabular}
\end{table*}

\begin{figure}
\includegraphics[width=9cm]{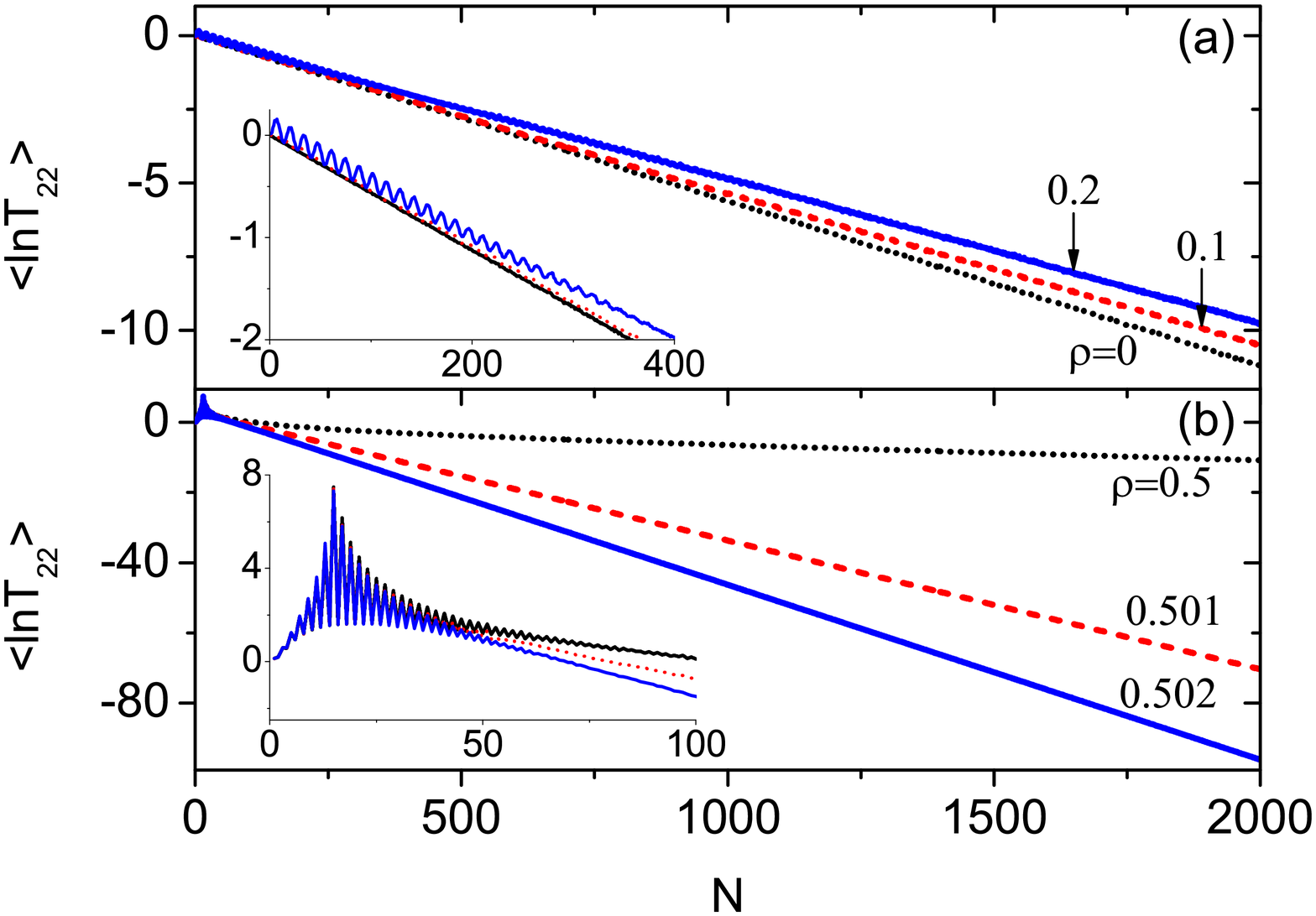}
\caption{Disorder-averaged logarithmic transmittance, $\langle \ln T_{22}\rangle$, plotted versus system size $N$, when $t_{v}=0.5$ and $\beta=0.5$, for (a) $\rho=0$, 0.1, 0.2 and (b) $\rho=0.5$, 0.501, 0.502. The insets show enlargements of the small-$N$ regions.}
\end{figure}

Next, we study the combined effects of the simultaneous presence of disorder and local $\mathcal{PT}$ symmetry on wave propagation. In this subsection, all disorder-averaged quantities are obtained by averaging over $30,000$ distinct disorder configurations.
In Fig.~8, we plot the disorder-averaged logarithmic transmittance $\langle \ln T_{22}\rangle$ as a function of the system size $N$, when the interchain coupling parameter $t_{v}$ is 0.5, for various values of the gain/loss strength $\rho$ in the presence of disorder with $\beta=0.5$. When $\rho$ is zero, we reproduce the result presented in Ref.~43, which shows that the disorder-averaged logarithmic transmittance in each channel decays linearly with the system size. When $\rho$ is non-zero but small, we find that the presence of a locally $\mathcal{PT}$-symmetric potential
suppresses the localization effect of disorder, giving rise to a slower decay of the logarithmic transmittance than for the case of $\rho=0$,
as shown in Fig.~8(a). In this parameter region, a periodic behavior reminiscent of the periodic oscillation in the clean system with $\rho<\rho_c$ can still be observed when $N$ is small, as seen in the inset of Fig.~8(a). The oscillatory behavior becomes more pronounced as $\rho$ increases
toward $\rho_c=0.5$, having a larger amplitude and persisting up to larger values of $N$. The oscillation amplitude decays gradually as $N$ increases to infinity.
The simultaneous occurrence of a periodic oscillation and a decaying behavior
is a consequence of the competition between disorder and $\mathcal{PT}$ symmetry in this regime.

We next consider the case where $\rho$ is larger than the critical value $\rho_{c}=t_v=0.5$. Similarly to the ordered case, $\langle \ln T_{22}\rangle$ takes a maximum value at a certain value of $N$, around which there appears
a rapid oscillatory behavior due to different dependencies on even and odd values of $N$, as illustrated in the inset of Fig.~8(b).
Contrary to the case where $\rho$ is small,
the decay rate is an increasing function of $\rho$ in this regime, as shown in Fig.~8(b).
This arises from the fact that in this regime, the decay of the transmission is achieved not only due to disorder, but also due to gain and loss.
By a linear fit of the data showing linear decay of the logarithmic transmittance,
we can extract the decay rate of transmittance corresponding to passive disordered ($1/\xi_{\beta}$), ordered locally $\mathcal{PT}$-symmetric ($1/\xi_{\rho}$) and disordered locally $\mathcal{PT}$-symmetric ($1/\xi_{\beta+\rho}$) cases, where
$\xi_\beta$ and $\xi_{\beta+\rho}$ are defined in a similar manner as in Eq.~(9), except that
$\ln T_{22}$ is replaced by $\langle\ln T_{22}\rangle$. Through a numerical analysis for different combinations of the parameters $\beta$ and $\rho$, we find that the approximation,
\begin{equation}
\frac{1}{\xi_{\beta+\rho}}\approx \frac{1}{\xi_{\beta}}+\frac{1}{\xi_{\rho}},
\end{equation}
is satisfied in the regime of weak disorder, as can be confirmed in Table I. In a purely 1D case, this relationship has been proposed in Ref.~13, where light transport in randomly layered optical media with global $\mathcal{PT}$ symmetry has been considered.

\begin{figure}
\includegraphics[width=9cm]{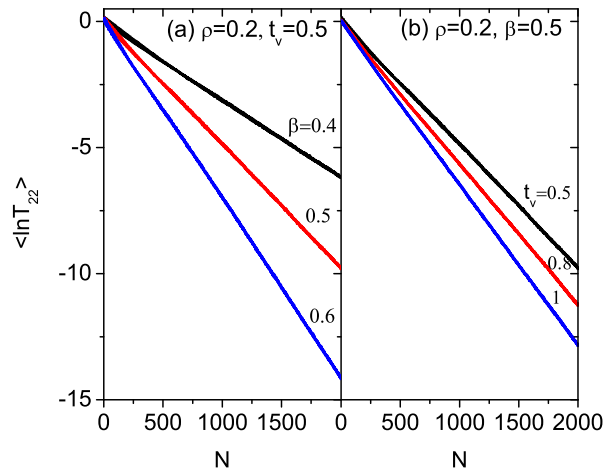}
\caption{$\langle \ln T_{22}\rangle$ plotted versus system size $N$ when $\rho=0.2$ for (a) $t_v=0.5$, $\beta=0.4$, 0.5, 0.6 and (b) $\beta=0.5$, $t_{v}=0.5$, 0.8, 1.}
\end{figure}

In Fig.~9, we plot $\langle \ln T_{22}\rangle$ versus $N$ when $\rho=0.2$ for different values of the disorder strength $\beta$ and of the interchain coupling parameter $t_{v}$. Similarly to the case of a real-valued potential, Anderson localization is enhanced when the strength of disorder increases as shown in Fig.~9(a). A similar behavior is also obtained when one fixes $\beta$ and varies $t_{v}$ as in Fig.~9(b). This is due to the fact that the role of $t_{v}$ is effectively equivalent to that of $\beta$ in the localization properties of ladder-structured lattices \cite{Ngu2}. We also find that the amplitude of the periodic oscillation of the logarithmic transmittance in the small-$N$ region decreases as either $\beta$ or $t_v$ increases. This again shows the competition between disorder and the gain/loss effect in this regime.

So far, all numerical results presented were obtained only for each individual channel and the energy of the incident wave was fixed to the band center value. The rest of the paper is devoted to the study of the localization properties in the entire system and over the whole energy spectrum. In Fig.~10, we plot the Lyapunov exponent $\gamma$, which is calculated using Eq.~(\ref{equation4}), as a function of the energy $E$, when $N=2000$, $\beta=0.5$, $t_{v}=0.5$ and $\rho=0$, 0.1, 0.2. In this regime where $\rho<t_v$, we find that the Lyapunov exponent decreases, and thus the localization length increases with increasing $\rho$ over the whole energy spectrum. When $\rho$ is zero, there are three values of energy for which the Lyapunov exponent spectrum exhibits a sharp peak or dip. This takes place precisely at the spectral positions $E=0$ and $E=\pm t_{v}$ as indicated by the arrows in Fig.~10. This is the phenomenon of anomalous localization, well-known in the study of passive disordered systems \cite{Ngu2,Czycholl,Kappus,Derrida,Izrailev}. We find that the anomaly at the band center manifests as a sharp peak, whereas the anomalies at $E=\pm t_{v}$ manifest as sharp dips. In other words, Anderson localization is anomalously enhanced at the band center and anomalously suppressed at $E=\pm t_{v}$. This is in contrast to the result of a recent study that all anomalies including the band center anomaly manifest as sharp dips \cite{Ngu2}.
In the ladder-shaped lattice model considered in Ref.~43,
the on-site random potentials $\epsilon_{n1}$ and $\epsilon_{n2}$ on chains 1 and 2 were
statistically uncorrelated, while they are correlated in the present study.
It has been reported that the band center anomaly depends strongly on the nature of disorder such as its correlation properties \cite{Titov,Car,Guox}. It is only associated with the disorder in the chains, therefore it remains unchanged as $\rho$ increases from zero.
In contrast, the positions of the side anomalies are shifted toward the band center when $\rho$ increases. By a numerical test for various pairs of the parameters $t_{v}$ and $\rho$, we have confirmed that these anomalies are dependent on the interchain coupling and the gain/loss strength
and always occur at $E=\pm\sqrt{t_{v}^2-\rho^2}$. The existence of this kind of anomalies is easily understood in terms of the $\pi$-coupling of the energy bands in the absence of disorder \cite{Ngu2,All}.

\begin{figure}
	\includegraphics[width=9cm]{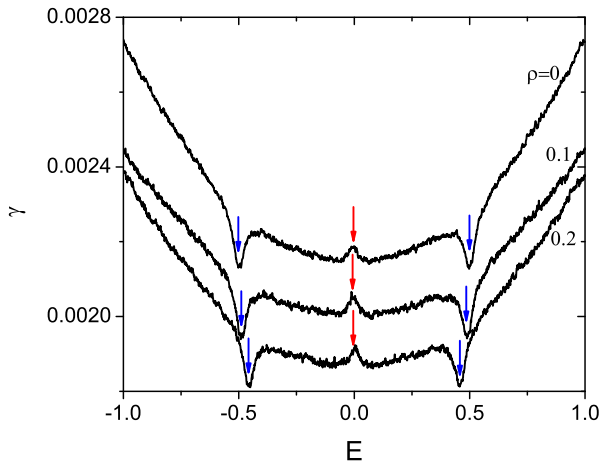}
	\caption{Lyapunov exponent $\gamma$ plotted versus energy $E$, when $N=2000$, $\beta=0.5$, $t_{v}=0.5$ and $\rho=0$, 0.1, 0.2. Anderson localization is suppressed overall in the presence of a locally $\mathcal{PT}$-symmetric potential with $\rho$ much smaller than $\rho_c$. Anomalous localization occurs precisely at the spectral positions $E=0$ (band center anomaly) and $E=\pm\sqrt{t_{v}^2-\rho^2}$ (side anomalies) as indicated by the arrows. Anderson localization at the band center is anomalously enhanced rather than suppressed unlike in the conventional cases.}
\end{figure}

\begin{figure}
\includegraphics[width=8.5cm]{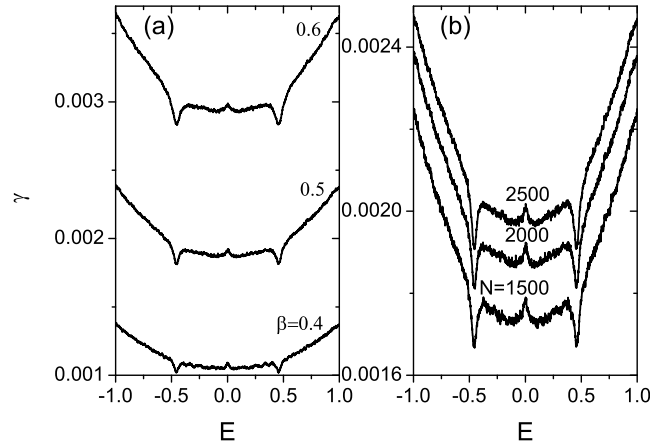}
\caption{Lyapunov exponent $\gamma$ plotted versus energy $E$ when (a) $N=2000$, $\rho=0.2$, $t_{v}=0.5$ and $\beta=0.4$, 0.5, 0.6, and when (b) $\rho=0.2$, $t_{v}=0.5$, $\beta=0.5$ and $N=1500$, 2000, 2500. The spectral positions at which the side anomalies occur do not change as either $\beta$ or $N$ varies.}
\end{figure}

Finally, in Fig.~11(a), we plot the Lyapunov exponent as a function of the energy $E$ when $N=2000$, $\rho=0.2$ and $t_{v}=0.5$, for several values of the disorder strength, $\beta=0.4$, 0.5, 0.6. Obviously, Anderson localization is enhanced with increasing $\beta$ over the whole energy spectrum. The spectral positions at which the side anomalies occur do not change as the disorder strength varies. This again confirms that these anomalies are only associated with the interchain coupling and the gain/loss parameter. In Fig.~11(b), we plot $\gamma$ versus $E$ when $\rho=0.2$, $t_{v}=0.5$ and $\beta=0.5$ for different system sizes $N=1500$, 2000, 2500. We find that all curves show a similar behavior. At any fixed value of $E$, the Lyapunov exponent increases with increasing $N$. The growth of $\gamma$ slows down for a larger $N$, and above a sufficiently large value of $N$, $\gamma$ approaches a saturation value. This result indicates that the system size $N = 2000$ is sufficiently large for the purpose of studying the localization properties in the system under consideration.

\section{Conclusion}

In this paper, we have presented a numerical study of the transport and localization properties of waves in ordered and disordered locally $\mathcal{PT}$-symmetric systems. We have employed the transfer matrix method developed in Ref.~37 to calculate the transmittance, the reflectance and the Lyapunov exponent. In the ordered case, we have found that when the gain/loss parameter $\rho$ is smaller than the interchain coupling parameter $t_{v}$, the transmittance and the reflectance are periodic functions of the system size. They have different dependencies on even and odd values of the system size. For a fixed system size, there appear transmission resonances in each individual channel at several values of the gain/loss strength. When $\rho$ is larger than $t_{v}$, we find that both the transmittance and the reflectance initially increase up to maximum values via oscillations and after that, the transmittance decays exponentially while the reflectance attains a saturation value as the system size increases. In addition, a singular behavior of the transmittance has also been shown. Next, as the disorder is introduced in the on-site potentials, these behaviors are changed substantially due to the interplay between disorder and the gain/loss effect. Specifically, when $\rho$ is smaller than $t_{v}$, we have found that the presence of locally $\mathcal{PT}$-symmetric potentials suppresses Anderson localization, as compared to the localization in the corresponding Hermitian system. In the case of $\rho>t_{v}$, it has been found that the localization becomes more pronounced at higher gain/loss strengths. Finally, the phenomenon of anomalous localization, which is well-known in passive disordered systems, has also been found to occur in locally $\mathcal{PT}$-symmetric systems. These anomalies occur precisely at the special spectral positions $E=0$ and $E=\pm\sqrt{t_{v}^2-\rho^2}$. The anomaly at the band center manifests as a sharp peak contrary to the conventional cases, whereas the anomalies at $E=\pm\sqrt{t_{v}^2-\rho^2}$ manifest as sharp dips.
We hope that the results presented here will be a useful contribution to the study of the combined effects of the simultaneous presence of disorder and $\mathcal{PT}$ symmetry on wave propagation in quasi-1D systems.

\acknowledgments
BPN is thankful to F. M. Izrailev and  F. A. B. F. de Moura for useful comments at the early stage of the present work.
This research is funded by Vietnam National Foundation for Science and Technology Development (NAFOSTED) under grant number 103.01-2014.10.
It is also supported by a National Research Foundation of Korea Grant (NRF-2015R1A2A2A01003494) funded by the Korean Government.

\newpage

\end{document}